\documentclass[twocolumn,twocolappendix,trackchanges]{aastex701}
\usepackage{amsmath}
\usepackage{microtype} 
\usepackage{ragged2e}  

\begin{document}

\title{OB stars identified in LAMOST Data Release 10}

\author{Guang Yang}
\affiliation{Department of Physics, Hebei Normal University, Shijiazhuang 050024, People's Republic of China}
\affiliation{Guo Shoujing Institute for Astronomy, Hebei Normal University, Shijiazhuang 050024, People's Republic of China}
\email{a18715911021@gmail.com}  

\author[orcid=0000-0001-5314-2924,gname=Zhicun, sname=Liu]{Zhicun Liu}
\affiliation{Department of Physics, Hebei Normal University, Shijiazhuang 050024, People's Republic of China}
\affiliation{Guo Shoujing Institute for Astronomy, Hebei Normal University, Shijiazhuang 050024, People's Republic of China}
\affiliation{Shijiazhuang Key Laboratory of Astronomy and Space Science}
\email[show]{liuzhicun@hebtu.edu.cn}

\author[orcid=0000-0003-2536-3142,gname=Xiao-Long,sname=Wang]{Xiao-Long Wang}
\affiliation{Department of Physics, Hebei Normal University, Shijiazhuang 050024, People's Republic of China}
\affiliation{Guo Shoujing Institute for Astronomy, Hebei Normal University, Shijiazhuang 050024, People's Republic of China}
\affiliation{Shijiazhuang Key Laboratory of Astronomy and Space Science}
\email{xlwang@hebtu.edu.cn}

\author[orcid=0000-0001-9989-9834,gname=Yanjun,sname=Guo]{Yanjun Guo}
\affiliation{Yunnan observatories, Chinese Academy of Sciences, P.O. Box 110, Kunming, 650011, People's Republic of China}
\affiliation{International Centre of Supernovae, Yunnan Key Laboratory, Kunming 650216, People's Republic of China}
\email{guoyanjun@ynao.ac.cn}

\author[orcid=0000-0003-1359-9908,gname=Wenyuan, sname=Cui]{Wenyuan Cui}
\affiliation{Department of Physics, Hebei Normal University, Shijiazhuang 050024, People's Republic of China}
\affiliation{Guo Shoujing Institute for Astronomy, Hebei Normal University, Shijiazhuang 050024, People's Republic of China}
\affiliation{Shijiazhuang Key Laboratory of Astronomy and Space Science}
\email[show]{cuiwenyuan@hebtu.edu.cn}




\correspondingauthor{Zhicun Liu and Wenyuan Cui}
\begin{abstract}

A large sample of OB stars plays an important role in studying the stellar parameters of massive stars, as well as the formation and evolution of the Milky Way. With the help of the Large Sky Area Multi-Object Fiber Spectroscopic Telescope (LAMOST) Data Release 10 (DR10), we are able to construct a large sample of OB stars with spectroscopic data. In this study, we identify 48,463 spectra of 34,550 OB stars from LAMOST DR10, based on the Hertzsprung–Russell (H-R) diagram constructed with \textit{Gaia} DR3 data and spectral line indices measured from LAMOST DR10 low-resolution spectra. Among these, 6907 OB stars are newly identified. We use the MKCLASS tool to derive the spectral subtypes of the OB sample. The spatial distribution of 25,287 OB stars and the Toomre diagram of 20,397 OB stars indicate that the majority of these stars are located in the Galactic disk. Based on their peculiar velocities, we identify 1960 runaway star candidates.

\end{abstract}



\section{Introduction} 

OB stars, defined here as O- and B-type stars with spectral types ranging from O2 to B9, have high effective temperatures and luminosities \citep{Morgan1973ARA&A..11...29M,Przybilla2012A&A...539A.143N}. OB stars are young stars and are not far from their birthplaces due to their short evolutionary timescales. These stars play an important role in the star formation and evolution of their host galaxies through mass-loss, ultraviolet radiation, and final supernova explosions \citep{Heger2003ApJ...591..288H,Gray2009ssc..book.....G}

OB stars, which are mainly located in the Galactic disk, are important probes for studying the structure and kinematics of the Milky Way. \citet{2019MNRAS.487.1400C} found the Galactic spiral structure exhibits flocculent patterns, based on 14,880 O- and early B-type stars with Gaia DR2 parallax uncertainties smaller than 20\%.
\citet{2021A&A...645L...8X} studied the Galactic spiral structure using 9750 O-B2 stars with distance accuracies better than 10\% selected from \citet{2014yCat....1.2023S}, and provided a detailed description of the local spiral structure. OB stars have also been used to trace the flare and warp of the Galactic disk \citep{2019ApJ...871..208L,2021ApJ...922...80Y}. \citet{2019AstL...45..331B} and \citet{2022AstL...48..243B} used OB stars with Gaia data to derive Galactic rotation parameters. \citet{2019ApJ...872L...1C} found that the Galactic disk shows a distinct ripple in radial and azimuthal velocities between 6 and 15\,kpc by analyzing the three-dimensional velocity distribution of about 12,000 OB stars.

OB runaway stars in the Milky Way have high space velocities or are located far from the Galactic plane. They originate via either the binary ejection mechanism or the dynamical ejection mechanism \citep{Silva2011MNRAS.411.2596S,McEvoy2017ApJ...842...32M,Liu2023MNRAS.519..995L}. In the binary ejection mechanism, OB runaway stars are produced via supernova explosions in binary systems, in which the massive primary explodes and ejects the lower-mass secondary as a high-velocity runaway star \citep{Zwicky1957moas.book.....Z,Blaauw1961BAN....15..265B}. In the dynamical ejection scenario, OB runaways result from gravitational interactions among stars in clusters or associations \citep{Poveda1967BOTT....4...86P}. The difference between the two ejection mechanisms is the distribution of ejection velocities and the chemical compositions of runaway stars \citep{Silva2011MNRAS.411.2596S,McEvoy2017ApJ...842...32M}. The origins and ejection mechanisms of runaway OB stars can be further investigated via studying their kinematics. 

Constructing a large sample of OB stars with multifaceted information helps us better understand the formation and evolution of the Milky Way. Spectroscopic and photometric datasets for large samples of OB stars have been presented by numerous studies \citep{Reed2003AJ....125.2531R,Sota2016ApJS..224....4M,MohrSmith2017MNRAS.465.1807M,Liu2019ApJS..241...32L,Liu2024ApJS..275...24L,Zari2021A&A...650A.112Z}. For instance, the intrinsic binary fraction, mass ratio, and orbital period distribution of OBA-type stars have been studied by \citet{Guo2022A&A...667A..44G}, based on data from the Large Sky Area Multi-Object Fiber Spectroscopic Telescope (LAMOST) Data Release 8 (DR8) medium-resolution survey. The nitrogen abundances and rotational velocities of B-type stars in the Large and Small Magellanic Clouds have been investigated \citep{Hunter2007A&A...466..277H,Hunter2008ApJ...676L..29H,2018A&A...615A.101D}. These studies suggest that rotation is not the only reason for the anomalous nitrogen abundances on the surfaces of B-type stars, and that magnetic fields, binary interactions, and blue loops also play important roles \citep{Dunstall2011A&A...536A..65D,McEvoy2015A&A...575A..70M}. 

Recently, the IACOB project presented a sample of approximately 500 Galactic O9-B9 stars, analyzed their rotational properties and found that there is a clear drop in the relative number of stars at effective temperatures of about 21000\,K \citep{deBurgos2023A&A...674A.212D,deBurgos2024A&A...687A.228D}. In addition, based on the distribution of upper limits on mass-loss rates for 116 Galactic luminous blue supergiants, \citet{deBurgos2024A&A...687L..16D} found no increase in mass-loss rates across the bistability region. Therefore, updating OB star samples is crucial for studying the formation and evolution of massive stars.

The paper is organized as follows: In Section \ref{sec:data}, we present the LAMOST DR10 dataset and the spectral line indices. Section \ref{sec:method} describes the process of identifying OB stars. The results and discussion are presented in Section \ref{sec:RaD}. Finally, a short summary of the work is provided in Section \ref{sec:Summary}.

\section{Data} \label{sec:data}

\subsection{The LAMOST data}

\begin{table}[!t]
    \footnotesize
    \centering
    \caption{The definition of line indices in this work}
    \begin{tabular}{lcc}
    \hline
    \hline
     Name&    Index Bandpass~(\AA) &Pseudocontinua~(\AA)\\
    
    \hline
    \ion{Ca}{2}~K &3927.700–3939.700 &3903.000–3923.000 4000.000–4020.000\\
    H$_\gamma$ & 4319.750–4363.500&4283.500–4319.750 4367.250–4419.750\\
    Fe\,4383&4370.375–4421.625&4360.375–4371.625 4444.125–4456.625\\
    Fe\,4531&4515.500–4560.500&4505.500–4515.500 4561.750–4580.500\\
    Fe\,4668&4635.250–4721.500&4612.750–4631.500 4744.000–4757.750\\
    Fe\,5015&4977.750–5054.000&4946.500–4977.750 5054.000–5065.250\\
    Fe\,5270&5245.650–5285.650&5233.150–5248.150 5285.650–5318.150\\
    Fe\,5335&5312.125–5352.125&5304.625–5315.875 5353.375–5363.375\\
    Fe\,5406&5387.500–5415.000&5376.250–5387.500 5415.000–5425.000\\
    Fe\,5709&5698.375–5722.125&5674.625–5698.375 5724.625–5738.375\\
    Fe\,5782&5778.375–5798.375&5767.125–5777.125 5799.625–5813.375\\
    \hline
    \end{tabular}
    \label{tab1:linedices}
\end{table}

The Large Sky Area Multi-Object Fiber Spectroscopic Telescope (LAMOST), also known as the Guo Shoujing Telescope, is a 4-meter-class quasi-meridian reflecting Schmidt telescope \citep{Cui2012RAA....12.1197C,Zhao2012RAA....12..723Z}. Its focal plane is equipped with 4,000 fibers, enabling the simultaneous acquisition of 4,000 spectra. The LAMOST DR10 data release contains a total of 11,544,238 low-resolution stellar spectra ($R\sim1800$) over the wavelength range of 3690–9100,\AA\footnote{These spectra correspond to all stars in LAMOST DR10v0, v1, and v2 with unique Obsid.}.

\begin{figure*}[!t]
    \centering
    \includegraphics[width=0.740\linewidth]{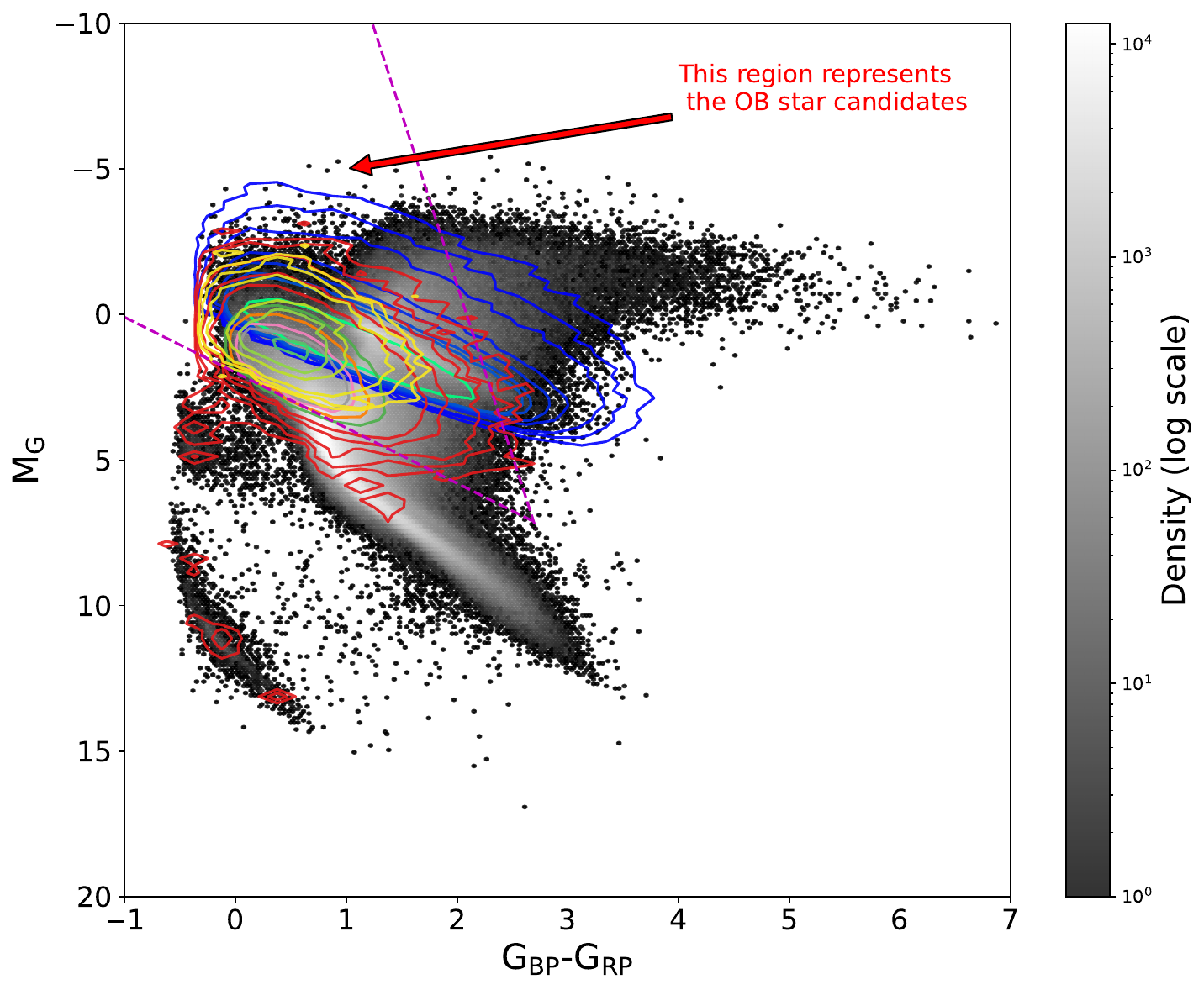}
    \caption{Distribution of 4,228,071 stars (gray dots) from LAMOST DR 10 on the \textit{Gaia} Hertzsprung-Russell diagram (M$_{\rm G}$ versus G$_{\rm BP}$-G$_{\rm RP}$). The blue, red, and yellow contours represent the distributions of OBA-type stars from \citet{Zari2021A&A...650A.112Z}, OBA-type stars from \citet{Xiang2022A&A...662A..66X}, and OB-type stars from \citet{Liu2024ApJS..275...24L}, respectively. The pink dashed lines mark the selection boundaries applied to remove contaminants, and the color bar indicates the stellar density.}
    \label{Fig01}
\end{figure*}

\subsection{line indices}

The spectral absorption lines in some stars are highly sensitive to their stellar parameters, such as effective temperature, surface gravity, and metallicity. The line indices of these absorption lines can be used to estimate the initial stellar parameters of the corresponding stars \citep{Liu2015RAA....15.1137L}. These indices, often referred to as equivalent widths (EWs), are defined as follows  \citep{Worthey1994ApJS...94..687W,Worthey1997ApJS..111..377W,Liu2015RAA....15.1137L}:

\begin{equation}\label{equation1}
    \rm EW = \int (1-\frac{F_\lambda}{F_C})
\end{equation}

In equation~\ref{equation1}, F$_\lambda$ and F$_{\rm C}$ represent the flux of the spectral line and the corresponding pseudo-continuum, respectively. F$_{\rm C}$ is estimated through linear interpolation of the flux in the “shoulder” region on either side of the line bandpass. Under this definition, the line index is expressed in units of \AA. According to previous studies, OB stars can be selected based on their distributions in the parameter spaces of specific spectral line indices \citep{Liu2019ApJS..241...32L,Liu2024ApJS..275...24L}. 
Therefore, we measure the EW of \ion{Ca}{2}\,K\,(3933 \AA), the EW of H$_\gamma$\,(4340\,\AA), and a set of Fe lines (EW$_{\rm Fe}$). The EW$_{\rm Fe}$ is the mean value of EWs of nine Fe lines at 4383, 4531, 4668, 5015, 5270, 5335, 5406, 5709, and 5782\,\AA. The definitions of all adopted line indices are listed in Table~\ref{tab1:linedices}.

\section{The process of identification of OB stars}\label{sec:method}

A number of OB star catalogs have been compiled from spectroscopic and photometric surveys \citep{MohrSmith2017MNRAS.465.1807M,Liu2019ApJS..241...32L,Zari2021A&A...650A.112Z,Liu2024ApJS..275...24L}. The LAMOST DR10 dataset contains 11,544,238 low-resolution stellar spectra, corresponding to 8,652,102 stars, among which a substantial number of OB stars are expected. To identify OB stars\footnote{In this work, we define OB stars as massive, young stars, excluding low-mass objects such as hot subdwarfs, white dwarfs (WDs), blue horizontal-branch stars, and post–asymptotic giant branch stars.}, we applied multiple screening criteria to remove contaminants. We selected spectra with a signal-to-noise ratio greater than 15 in the g-band (S/N$_{\rm g}$) as our sample, obtaining 6,778,253 spectra. In addition, 30,686 spectra with S/N$_{\mathrm g}$ = $-9999$ were retained. This brings the total initial sample to 6,808,939 (6,778,253+30,686) spectra, corresponding to 5,217,192 unique stars.

\begin{figure*}[htp!]
    \centering
    \includegraphics[width=1.0\linewidth]{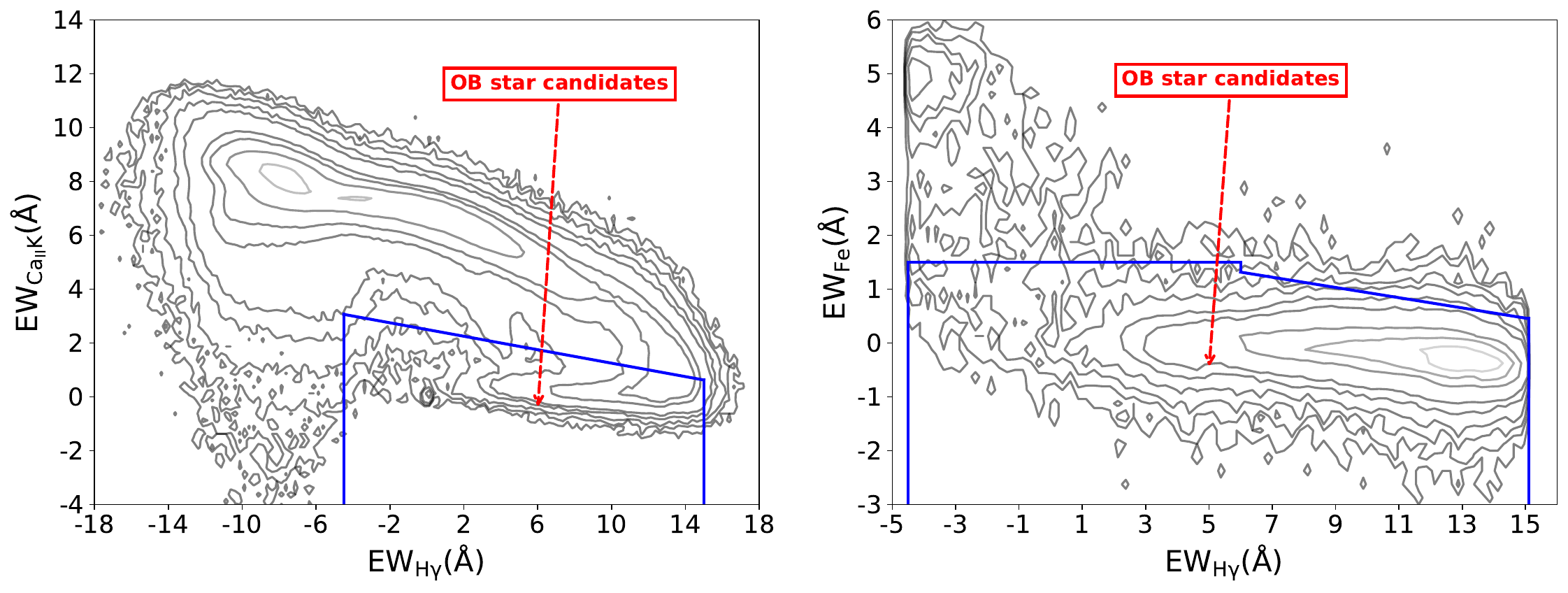}
    \caption{Left panel: distribution of 3,426,392 spectra of 2,621,576 OB star candidates in the EW$_{\rm Ca_{\rm II}K}$ versus EW$_{\rm H_\gamma}$ plane. The blue solid lines represent the cuts used to select OB star candidates, following Equation \ref{equation2}. Right panel: distribution of 171,836 spectra of 133,550 OB star candidates in the EW$_{\rm Fe}$ versus EW$_{\rm H_\gamma}$ plane. The blue solid lines indicate the cuts used to remove contaminants, following Equation \ref{equation3}. The selected OB star candidates are highlighted in red.}
    \label{Fig02}
\end{figure*}

\subsection{Selecting OB star candidates in the Hertzsprung–Russell diagram}

Different types of stars can be identified based on their positions in the Hertzsprung–Russell (H-R) diagram. Using the accurate parallaxes and photometry from the third data release (DR3) of \textit{Gaia} \citep{Gaia2023A&A...674A...1G}, we can build an H-R diagram for a large sample of stars. By cross-matching the 5,217,192 stars from LAMOST with \textit{Gaia} DR3 data, we obtained 4,228,071 common stars that have parallax ($\varpi$) measurements, photometric data, and satisfy the following quality criteria: a renormalised unit weight error (RUWE)$\leq$1.4 and $\frac{\rm \sigma_\varpi}{\rm \varpi}\leq$20\%. The absolute magnitude (M$_{\rm G}$) is calculated using the relation M$_{\rm G}$ = $G$+5+5log$_{10}$ ($\varpi$/1000), where $\varpi$ is the parallax in milliarcseconds \citep{Gaia2018A&A...616A..10G}.

To define a suitable boundary in the H-R diagram for identifying OB stars, we collected the known OBA-type stars from \citet{Zari2021A&A...650A.112Z}, \citet{Xiang2022A&A...662A..66X}, and \citet{Liu2024ApJS..275...24L} and computed their absolute magnitudes. In Figure~\ref{Fig01}, we map the 4,228,071 stars in LAMOST DR10 together with the OBA-type stars from the aforementioned references on the M$_{\rm G}$ versus G$_{\rm BP}$-G$_{\rm RP}$ plane. The OBA-type stars were processed using the same procedure as applied to the OB stars. The known OB stars are located in the upper-left region of the H-R diagram, owing to their high effective temperatures and luminosities, whereas sdBs and WDs are situated in the lower-left region \citep{Lei2018ApJ...868...70L}. To maximize the number of OB stars in our study, we empirically defined the selection region bounded by the pink dashed lines to include most known OB stars. The selection criteria are as follows:

\begin{equation}\label{equation11}
\begin{aligned}
    \rm M_G<1.9*(G_{BP}-G_{RP})+2.0 \\
    \rm M_G>11.75*(G_{BP}-G_{RP})-24.9
\end{aligned}
\end{equation} 

Applying the above criteria, we identified 2,163,308 spectra of 1,633,473 OB star candidates. In addition, 1,264,573 spectra of 989,142 stars were excluded due to a lack of \textit{Gaia} photometry in the blue and red bands, unavailable parallax measurements, RUWE values greater than 1.4, or excessively large parallax uncertainties \footnote{Among these 989,142 stars, 25,820 lack Gaia astrometric data, 54,740 are missing Gmag, BP, RP, or parallax measurements, 615,898 have RUWE$\geq$1.4, and 296,684 have relative parallax uncertainties greater than 20\%.}. Thus, the total number of OB star candidate spectra is 3,427,881 (2,163,308+1,264,573), corresponding to 2,622,615 unique stars.

\subsection{Selecting the OB stars candidates in the line indices' space}

\begin{figure*}[!t]
    \centering
    \includegraphics[width=1.0\linewidth]{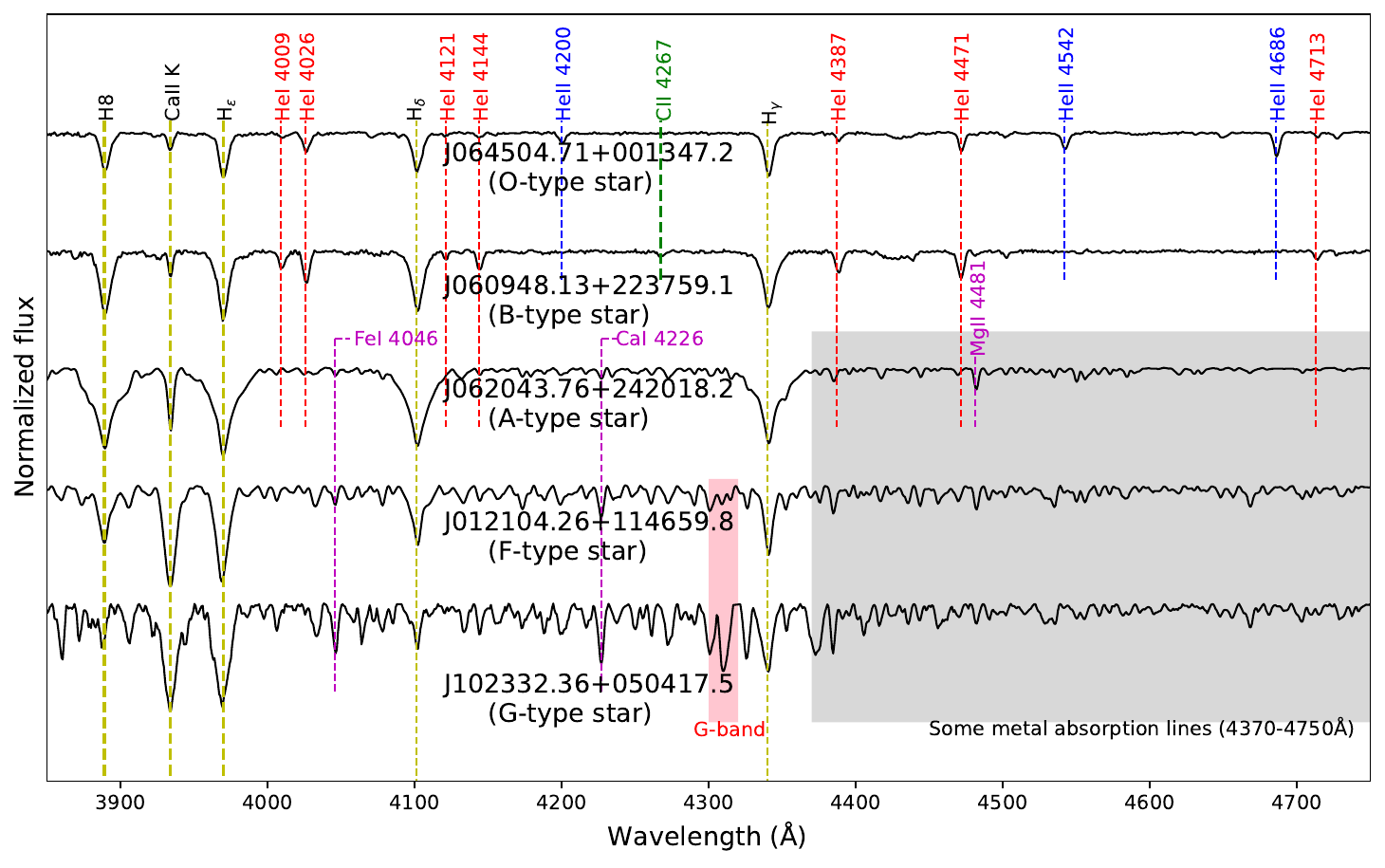}
    \caption{From top to bottom, the figure shows the blue-region spectra of five stars with spectral types O, B, A, F, and G from LAMOST. Their main characteristic spectral absorption lines are labeled. For O- and B-type stars, these include \ion{He}{2} 4200, 4541, 4686\,\AA, \ion{He}{1} 4009, 4026, 4121, 4144, 4387, 4471, 4713\,\AA\, ,and \ion{C}{2} 4267\,\AA. For A-, F-, and G-type stars, the marked features include \ion{Ca}{2}\,K, \ion{Fe}{1} 4046\,\AA, \ion{Ca}{1} 4226\,\AA, \ion{Mg}{2} 4481\,\AA, and the G-band.}
    \label{Fig03}
\end{figure*}

Following the method of \citet{Liu2019ApJS..241...32L,Liu2024ApJS..275...24L}, OB stars can be selected based on their positions in the EW$_{\rm H_{\rm \gamma}}$ versus EW$_{\rm Ca_{\rm II}K}$ and EW$_{\rm H_\gamma}$ versus EW$_{\rm Fe}$ planes. In the left panel of Figure~\ref{Fig02}, we plot 3,426,392 of the 3,427,881 initial OB star candidates \footnote{Among the 3,427,881 OB star candidate spectra, 1489 have no measurable values for EW$_{\rm Ca_{\rm II}K}$, EW$_{\rm H_\gamma}$, or EW$_{\rm Fe}$ due to poor flux quality in the corresponding spectral line regions.}  in the EW$_{\rm H_{\rm \gamma}}$ versus EW$_{\rm Ca_{\rm II}K}$ plane and apply the following empirical criteria to select OB stars:

\begin{equation}\label{equation2}
\begin{aligned}
    \rm EW_{\rm CaK}<2.5-EW_{\rm H\gamma}/8 \\
    -4.5\,\leq\,\rm EW_{\rm H\gamma}\,\leq\,15
\end{aligned}
\end{equation} 

After applying criteria~\ref{equation2}, we obtained 171,836 spectra of 133,550 OB star candidates. In the right panel of Figure \ref{Fig02}, we plot the above OB star candidates in the EW$_{\rm H_\gamma}$ versus EW$_{\rm Fe}$ plane and use the following empirical criteria to further select OB star candidates.

\begin{equation}\label{equation3}
\begin{aligned}
    \rm EW_{\rm Fe}<1.5\,and\,EW_{\rm H\gamma}<6 \\
    \rm EW_{\rm H\gamma}>6 \, and\, EW_{\rm Fe}<-0.095*EW_{\rm H\gamma}+1.316 \,
\end{aligned}
\end{equation} 

After applying criteria~\ref{equation3}, the number of OB star candidate spectra is reduced to 157,259. In addition, there are 1489 spectra without measured values of EW$_{\rm Ca_{\rm II}K}$, EW$_{\rm H_\gamma}$, and EW$_{\rm Fe}$ because their spectra have bad fluxes in the relevant wavelength regions. Thus, our final OB star candidate sample comprises 158,748 spectra (157,259+1489) of 123,664 stars (122,214+1450) .

\begin{table*}[htp!]
    \footnotesize
    \centering
    \caption{Information of OB stars identified in LAMOST DR10.}
    \begin{tabular}{lccccccc}
    \hline
    \hline
    Parameter &Unit& Star 1 & Star 2 & Star 3 & \nodata & \nodata &\nodata \\
    \hline
    Obsid (LAMOST)& &269516124	&269515157	&66604134&\nodata&\nodata&\nodata \\
    Designation (LAMOST)& & J000005.68+580.2&J000006.61+581041.6	&J000009.65+115333.9	&\nodata&\nodata&\nodata  \\
    \textit{Gaia}  DR3 Name& & 422764248104468	&422728621360546560	&2765768108934958080	&\nodata&\nodata&\nodata  \\
    RA (LAMOST)&deg & 0.0236781	&0.027556	&0.040234	&\nodata&\nodata&\nodata  \\
    Dec (LAMOST)&deg & 58.5945138	&58.17823	&11.892757	&\nodata&\nodata&\nodata \\
    S/N$_g$ (LAMOST)& & 78.56	&241.74	&62.29	&\nodata&\nodata&\nodata  \\
    Plx  (\textit{Gaia}  DR3)&mas&0.865	&1.2154	&0.2168	&\nodata&\nodata&\nodata \\
    e\_Plx (\textit{Gaia}  DR3)&mas& 0.0125	&0.0161	&0.0385	&\nodata&\nodata&\nodata \\
    Gmag (\textit{Gaia}  DR3)& mag&12.360595	&11.447089	&14.96295	& \nodata&\nodata&\nodata \\
    BPmag (\textit{Gaia}  DR3)&mag&12.614079	&11.618545	&15.001565	&\nodata&\nodata&\nodata  \\
    RPmag (\textit{Gaia}  DR3)& mag& 11.945554	&11.139322	&14.879053	&\nodata&\nodata&\nodata \\
    rpgeo$^a$ & pc&1132.41418	&787.626221	&3814.7124	&\nodata&\nodata&\nodata  \\
    b\_rpgeo\_x$^b$ & pc &1117.88428	&778.381042	&3316.53979	&\nodata&\nodata&\nodata \\
    B\_rpgeo\_xa$^c$& pc &1146.52551	&797.965088	&4457.00586	& \nodata&\nodata&\nodata \\
    SIMBAD Name&  & &TYC 3664-324-1&2MASS J00000963+1153341	& \nodata&\nodata&\nodata \\
    sp\_type (SIMBAD)&  & &&B9.3	
    &\nodata&\nodata&\nodata \\
    sp\_type (MKCLASS)& &A8 mA0 V metal-weak &A0 IV &A9 mA0 V Lam Boo&\nodata&\nodata&\nodata \\
    Comment$^d$&  & 1020		&1020&1020&\nodata&\nodata&\nodata  \\
    GroupID$^e$&&  &&	713&\nodata&\nodata&\nodata  \\
    GroupSize$^f$ & &&&2&\nodata&\nodata&\nodata  \\
    \hline
    \end{tabular}
    \vspace{0.3em}
    {\footnotesize 
    \raggedright \\
    $^a$ The rpgeo represents the stellar photogeometric distance from \citet{Baj2021AJ....161..147B}.\\
    $^b$ and $^c$ The b\_rpgeo\_x and B\_rpgeo\_xa represent the value at 16\% and 84\% of the distance from \citet{Baj2021AJ....161..147B}, respectively.\\
    $^d$ The number represents which version of the LAMOST survey the spectrum comes from, such as: 1020 represents LAMOST DR10V2.0.\\
    $^e$ The ID number of the star for which was observed many times.\\
    $^f$ The number of exposures for the star.\\
    Note: This table is available in its entirety in a machine-readable form in the online journal.\\}
    \label{Table2}
\end{table*}

To maximize the completeness of our OB star sample, we visually inspected the 158,748 spectra of the 123,664 OB candidates based on the known spectral characteristics of OB stars \citep{Gray2009ssc..book.....G,Liu2019ApJS..241...32L}. In Figure~\ref{Fig03}, we show the blue-region spectra (3850-4750\,\AA) of five stars with spectral types O, B, A, F, and G from LAMOST DR10. Compared to A-, F-, and G-type stars, the main characteristics of OB stars are the presence of \ion{He}{2} and \ion{He}{1} absorption lines, along with weak metal absorption lines. Stars whose spectra show \ion{He}{2} 4200, 4541, 4686\,\AA absorptions, together with some weak \ion{He}{1} lines (such as \ion{He}{1} 4026, 4387, 4471\,\AA ) are classified as O-type stars. Stars whose spectra exhibit \ion{He}{1} 4009, 4026, 4387, 4471\,\AA, and \ion{Mg}{2}\,4481\,\AA absorption lines, but lack \ion{He}{2} 4200, 4541, 4686\,\AA absorption lines, are classified as B-type stars. Following the criteria described above, we identified 47,789 spectra of 33,876 OB stars. By cross-matching with the OB star catalog from \citet{Liu2024ApJS..275...24L}, we find that 614 known OB stars are missing from our initial selection. The reasons are as follows:

(i) 159 stars at the boundary of criteria \ref{equation2} and \ref{equation3} are missed because the different S/N of the spectra in different LAMOST data releases caused slight changes in the spectral indices. 

(ii) 229 stars are missed because they do not satisfy criterion \ref{equation11} in the H-R diagram.

(iii) 7 stars with spectral S/N$_g\leq$15 in LAMOST DR10 are missed. 

(iv) 112 stars with spectral S/N$_g\geq$15 from LAMOST DR5 to DR9 are not included in DR10.

Therefore, we add these 614 OB stars to our final sample. We also include the 60 stars identified during the recovery of OB stars verification procedure described in Section~\ref{sec:Completeness}. The final sample contains 48,463 spectra, corresponding to 34,550 OB stars. The steps for identifying OB stars are summarized in Figure~\ref{figureA1}. By cross-matching our results with the OB star catalog from \citet{Liu2024ApJS..275...24L}, we find that 6907 OB stars are newly identified. The basic information for the final OB star sample is provided in Table~\ref{Table2}.

\subsection{Recovery of OB stars}\label{sec:Completeness}

To test the recovery of method for identifying OB stars in LAMOST DR10, we randomly selected 20,000 stars with spectral S/N$_{\rm g}$$\geq$15, and manually inspected their spectra. We identified 113 OB stars. By cross-matching with our OB star catalog, we found that all 113 are included in our sample. We also collected stars classified as O and B spectral types from the initial LAMOST DR10 catalog and obtained 20,161 spectra of 17,100 stars. Cross-matching with the 34,490 OB stars, we obtained 7522 OB stars. For the remaining 9,753 stars, we visually inspected their spectra and identified 60 OB stars. Of these 60 OB stars, 54 were omitted because the spectral S/N$_{\rm g}$ was below 15, 3 were excluded by the criterion in equation~\ref{equation1}, and 3 were omitted because of the incorrect values of EW$_{\rm Ca_{\rm II}K}$, EW$_{\rm H_\gamma}$, and EW$_{\rm Fe}$ caused by their bad spectra. Overall, 7522 out of 7528 OB stars (99.92\%) with spectral S/N$_g\geq$15 in the initial LAMOST DR10 catalog have been identified. 

Based on the above analysis, we suggest that OB stars with spectral S/N$_{\rm g}$$\geq$15 in the initial LAMOST DR10 catalog may have been identified as completely as possible using our method. We note, however, that because LAMOST can only observe a subset of targets, the recovery here—defined by the observational selection function—measures only the OB stars recovered from the observed spectra, which is distinct from the completeness with respect to the true underlying OB star population.

\begin{figure*}[!t]
    \centering
    \includegraphics[width=1.0\linewidth]{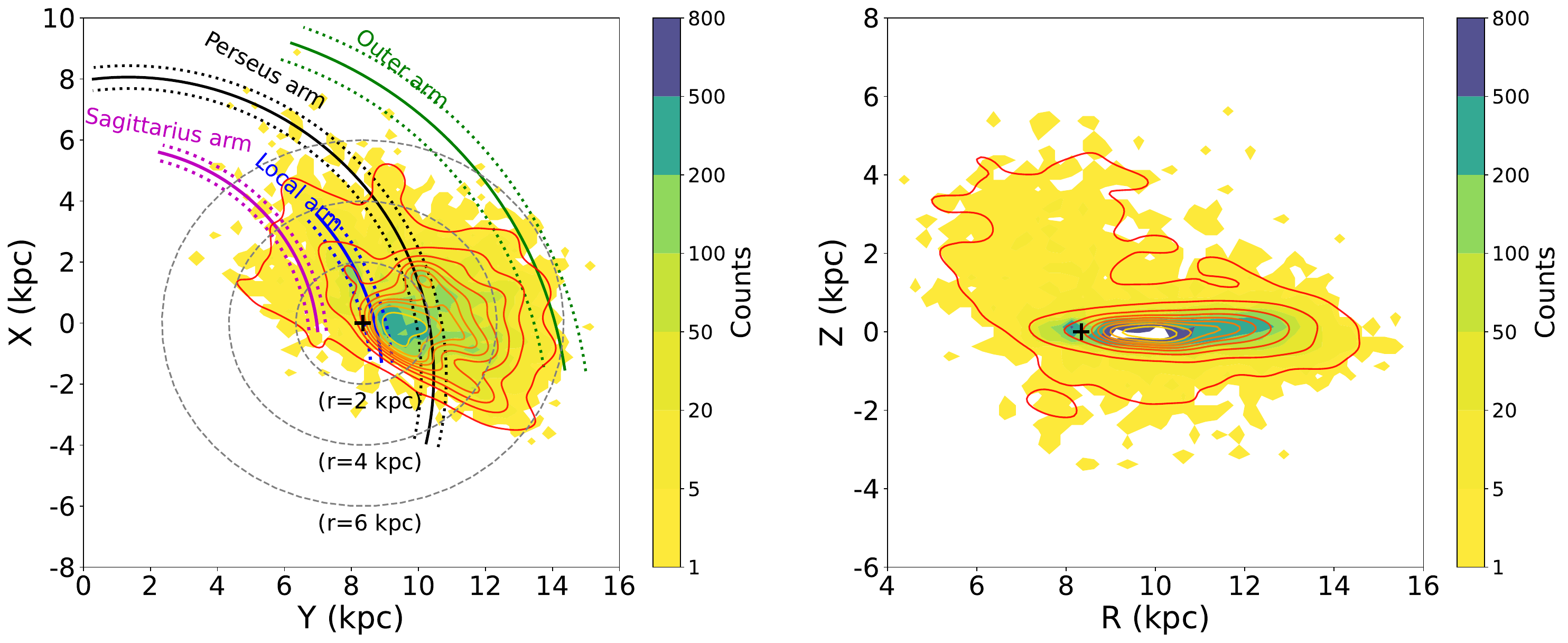}
    \caption{Spatial distributions of OB stars identified in LAMOST DR10. The background image (unfilled contours) represents 25,287 OB stars with RUWE$\leq$1.4 and $\sigma_\omega$/$\omega\leq$30\%, while the red contours represent the 5048 newly identified sources. Left panel: distribution in the X–Y plane of the Galactocentric Cartesian coordinate system.  Right panel: distribution in the R–Z plane of the Galactocentric Cartesian coordinate system. The position of the Sun is marked by the black cross (X=8.34\,kpc, Y=0\,kpc, Z=0\,kpc) ,and the data used to plot the Galactic spiral arms are both derived from \citet{Reid2014ApJ...783..130R}.
    The black dashed rings in the left panel delineate constant distances from the Sun in steps of 2\,kpc. The color bar corresponds to the number density of stars.}
    \label{Figure05}
\end{figure*}

\section{Results and Discussion} \label{sec:RaD}

\subsection{MKCLASS classification results}

To obtain the spectral types of our OB star sample, we used the automatic classification code MKCLASS \citep{2014AJ....147...80G,2016AJ....151...13G}. \citet{2014AJ....147...80G} developed MKCLASS code to classify blue–violet spectra within the MK system. The code works by minimizing the $\chi^2$ between the spectra of program stars and those of MK standard stars \citep{1943assw.book.....M}.
MKCLASS provides classification results with flags "excellent", "verygood", "good", "fair", and "poor". It contains two standard libraries, $libnor$18 and $libnor$36. $libnor36$ contains rectified blue–violet (3800–5600\,\AA) spectra of MK standard stars, obtained with the Dark Sky Observatory at a resolving power of $R\sim1100$. $libr18$ is built from flux-calibrated spectra covering 3800-4600\,\AA\,at a resolving power of $R\sim2200$.

LAMOST low-resolution spectra have a resolving power of $R\sim1800$ corresponding to a full width at half maximum of 2.8\,\AA. Thus, we convolved the library $libnor$18 from 1.8 to 2.8\,\AA and defined the resulting library as $libnor18\_{28}$\citep{Liu2019ApJS..241...32L,Liu2024ApJS..275...24L}. Then, we submitted the 48,463 spectra of 34,550 OB stars to MKCLASS and obtained their spectral types. For stars with multiple observations, we used the highest-S/N spectrum to obtain the spectral type to facilitate the analysis. The MKCLASS classification results are also listed in Table \ref{Table2}. Among the 34,550 OB-type stars, the subclassification results are as follows: 24,638 stars have both definite spectral types and luminosity classes; 101 stars have spectral types but no luminosity classes; 580 stars remain unclassified owing to low S/N or a lack of emission lines; and the remaining 8950 stars exhibit ambiguous spectral types based on different characteristic lines. The spectral types of 22,960 of 24,638 OB-type stars are assigned an MKCLASS quality class of "good" or higher. The classification results of 34,550 OB type stars are further discussed in Appendix \ref{MKOB}.

\begin{figure*}[!tp]
    \centering
    \includegraphics[width=1.0\linewidth]{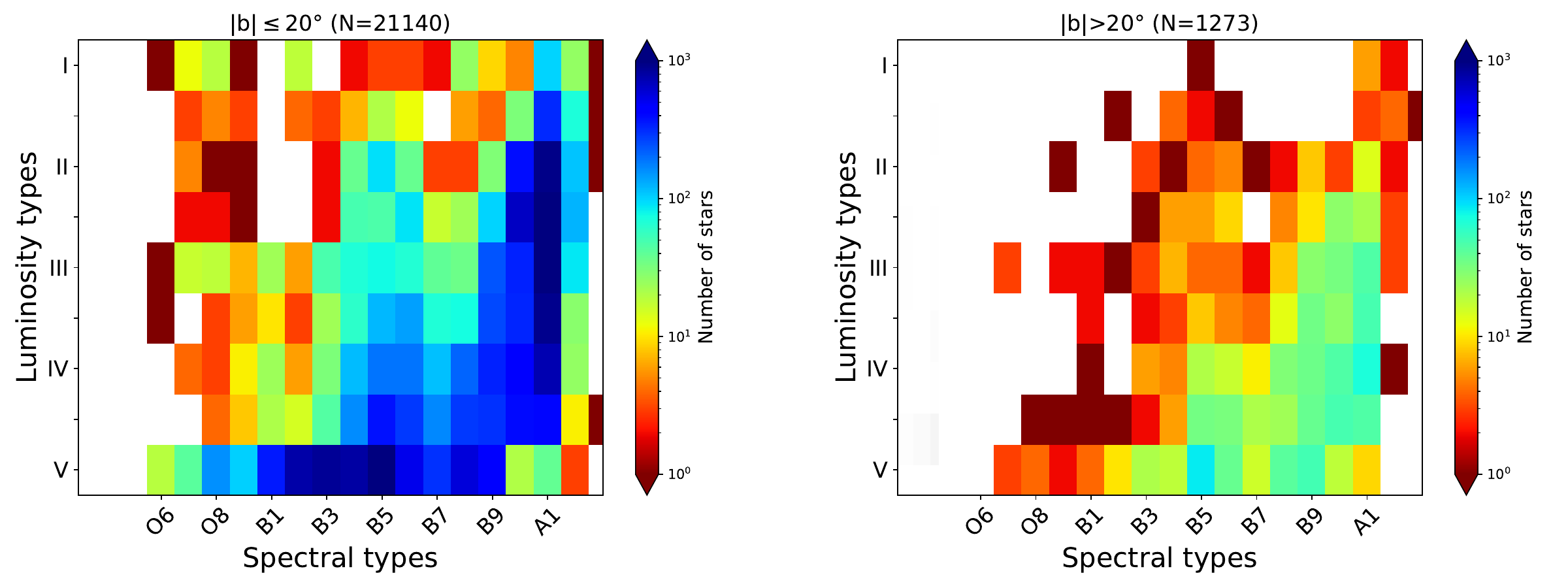}
    \caption{Left panel: distribution of the identified OB stars with Galactic latitudes $|b|\leq20^\circ$ in the luminosity class versus spectral subtype plane. Right panel: distribution of the identified OB stars with Galactic latitudes $|b|>20^\circ$ in the luminosity class versus spectral subtype plane. The colorbar represents the number of stars in each bin on a logarithmic scale.}
    \label{Figure050}
\end{figure*}

\begin{figure*}[!htp]
    \centering
    \includegraphics[width=0.89\linewidth]{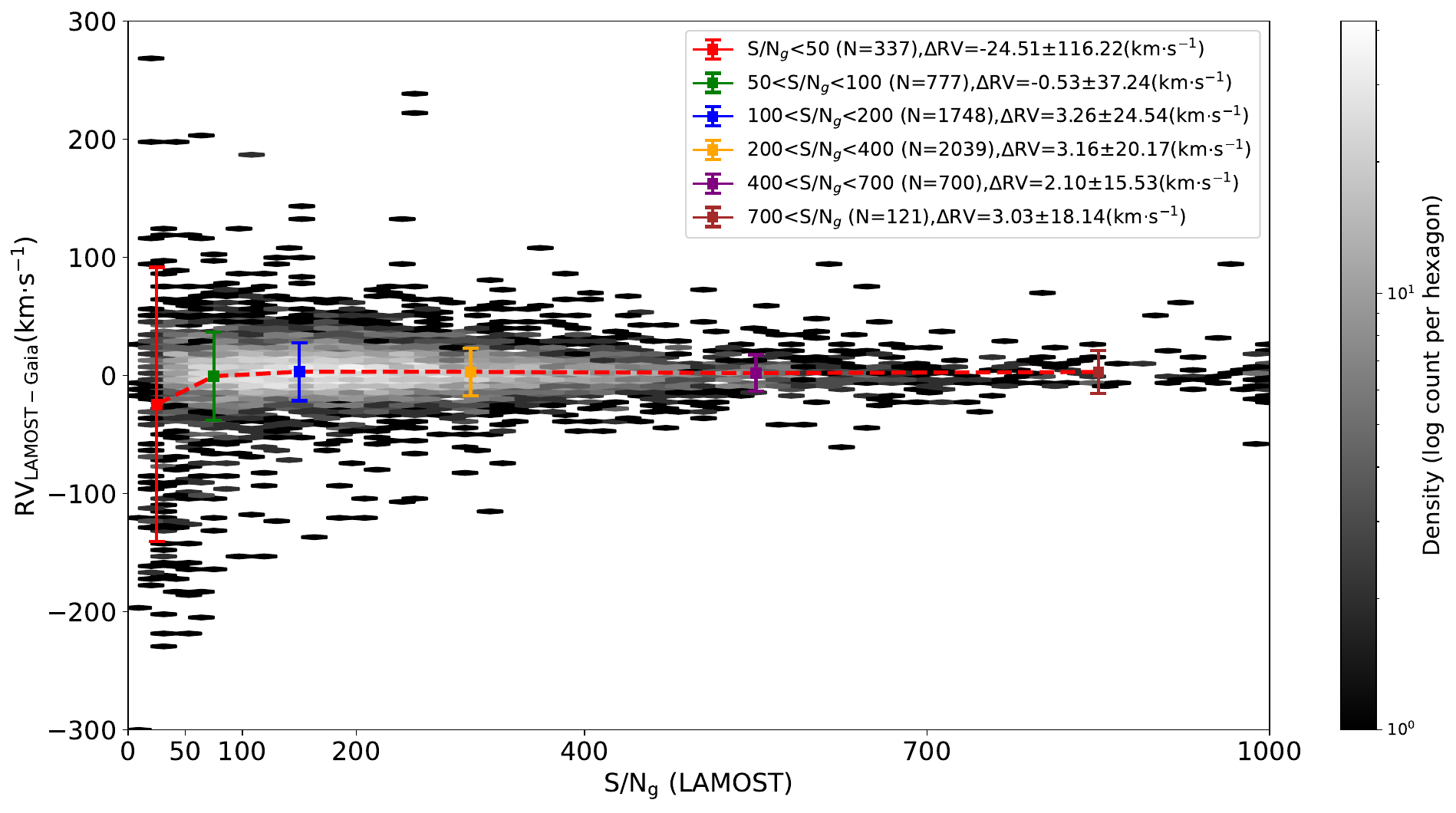}
    \caption{The radial velocity difference between LAMOST and Gaia (RV$_{\rm LAMOST-Gaia}$) as a functions of S/N$_{\rm g}$ value in LAMOST spectra for 5727 common stars. The color bar corresponds to the stellar density. The sample is divided into five bins according to S/N$_{\rm g}$: $<$50, 50–100, 100–200, 200–400, and 400–1000. The error bars in different colors denote the mean and standard deviation of RV$_{\rm LAMOST-Gaia}$ ($\Delta$RV) for all stars within each bin. The number of stars in each bin is indicated.}
    \label{Figure055}
\end{figure*}

\subsection{The spatial distribution of OB stars}
Based on the \textit{Gaia} DR3 data, we select 25,287 OB stars that satisfy the criteria RUWE $\leq$1.4 and $\sigma_\omega$/ $\omega\leq$30\%. Their Galactocentric coordinates (X, Y, Z) were calculated using photogeometric distances from \citet{Baj2021AJ....161..147B} and the solar position from \citet{Reid2014ApJ...783..130R} (R$_\odot$ = 8.34 kpc). Among these, 5048 OB stars belong to the 6907 newly identified ones. 

In Figure~\ref{Figure05}, we display the distribution of 25,287 OB stars and the 5048 newly identified ones in the Galactic X-Y (left panel) and the R-Z (right panel) planes, where the Galactic spiral arm structure from \citet{Reid2014ApJ...783..130R} is also marked. From the left panel of Figure \ref{Figure05}, it is seen that most OB stars are located between the Local arm and the Perseus arm. Our OB stars can also trace the Outer arm, and the spatial distribution of OB stars suggests that there are star formation regions between the Galactic spiral arms. Compared with the previous results, the newly identified OB stars (represented by the red kernel density estimation) show a similar distribution. The right panel of Figure~\ref{Figure05} further confirms that these newly identified OB stars are also predominantly located within the Galactic disk. However, it is important to note that the recovery analysis in Section \ref{sec:Completeness} quantifies identification reliability relative to the test sample, not the absolute completeness of the Galactic OB star population. Our spatial distributions are subject to selection biases inherent to the source catalog (e.g., magnitude limits and sky coverage). Consequently, our results should be interpreted as tracing the detected and validated sample rather than providing an unbiased depiction of the entire underlying OB star population.

\begin{figure}[!htp]
    \centering    
    \includegraphics[width=1.0\linewidth]{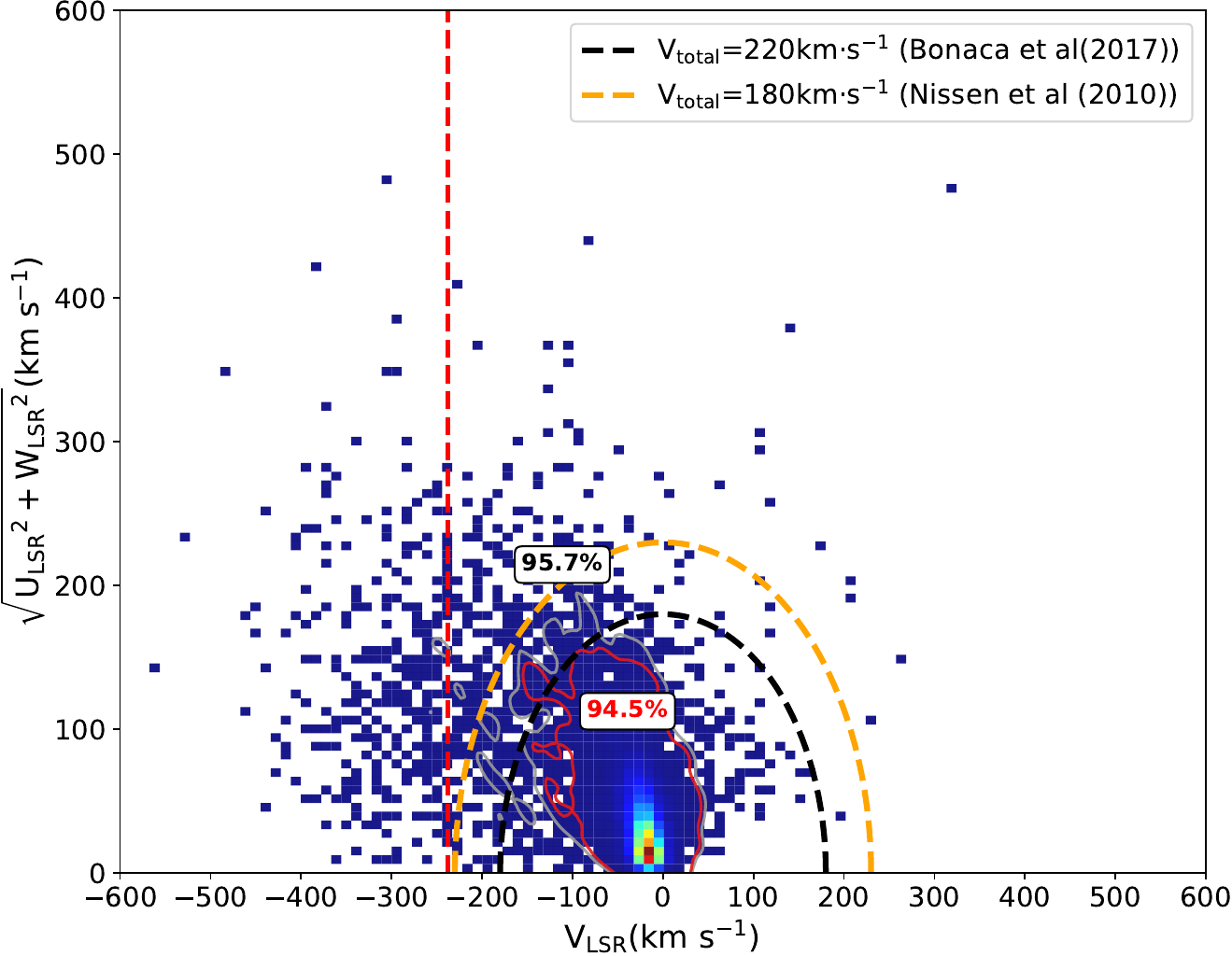}
    \caption{Toomre diagram for 20,397 OB stars that satisfy RUWE $\leq$1.4, $\sigma_\omega$/$\omega\leq$30\%, and have LAMOST or \textit{Gaia} radial velocities. The orange and black dashed semicircles correspond to V$_ {\rm total}$=220\,km$\cdot$s$^{-1}$ and V$_{\rm  total}$=180\,km$\cdot$s$^{-1}$, respectively, where V$_{\rm total}$=(V$_{\rm LSR}^2$+U$_{\rm LSR}^2$+W$_{\rm LSR}^2$)$^{1/2}$. These velocity thresholds are used to separate halo stars from disk stars. The unfilled contours contain 94.5\% and 95.7\% of the sample, respectively.}
    \label{Figure06}
\end{figure}

In Figure~\ref{Figure050}, we show the distribution of 22,413 OB stars with MKCLASS quality flags of “good” or better and spectral types ranging from O6 to A2 in the plane of spectral subtype and luminosity class, for Galactic latitudes less than 20\,$^\circ$ (left panel) and greater than 20\,$^\circ$ (right panel). The figure shows that 95.4\% of the stars are located at  low Galactic latitudes. Compared with stars at low Galactic latitudes, stars located at high Galactic latitudes are mainly B3–A1 stars with luminosity classes IV–V. These high-latitude OB stars are also ideal tracers for investigating the origins of runaway stars \citep{2004AJ....128.2474M,2006AJ....131.3047M,Liu2023MNRAS.519..995L}.

\subsection{The kinematics of OB stars}

To further explore the spatial distribution of OB stars in the LAMOST survey, we selected 25,168 OB stars with radial velocities from the LAMOST survey, geometric distances from \citet{Baj2021AJ....161..147B}, RUWE$\leq$1.4, and proper motions (‘pmra’, ‘pmdec’) from the \textit{Gaia} DR3 database \citep{Gaia2023A&A...674A...1G}. Previous studies have indicated that the radial velocities obtained from LAMOST spectra have systematic deviations. We cross-matched our 25,168 OB stars with \textit{Gaia} data and obtained 5727 common OB stars with radial velocity measurements. In Figure~\ref{Figure055}, we present the radial velocity difference (RV$_{\rm LAMOST-Gaia}$) between LAMOST and Gaia as a function of the S/N$_{\rm g}$ value in the LAMOST spectra. When S/N$_{\rm g}$ is below 50, the RV$_{\rm LAMOST-Gaia}$ is relatively large, while when S/N$_{\rm g}$ exceeds 50, the RV$_{\rm LAMOST-Gaia}$ exhibits systematic variations. Therefore, we used the 5385 stars with S/N$_{\rm g}$$\geq$50 to calculate the difference between RV$_{\rm LAMOST}$ and RV$_{\rm Gaia}$, and found that the LAMOST radial velocities are systematically larger than the Gaia values by $2.52$ km s$^{-1}$. Consequently, for the 14,670 OB stars that have only LAMOST radial velocities and spectral S/N$_{\rm g}\geq50$, we applied a correction of $-$2.52\,km$\cdot$s$^{-1}$. For the 5,727 common OB stars, we adopted the Gaia radial velocities. The Galactic space velocities (U$_{\rm LSR}$, V$_{\rm LSR}$, and W$_{\rm LSR}$) for the 20,397 ($=14,670 + 5,727$) stars were then calculated using the parameters described above. 

\begin{figure*}[!htp]
    \centering
    \includegraphics[width=1.0\linewidth]{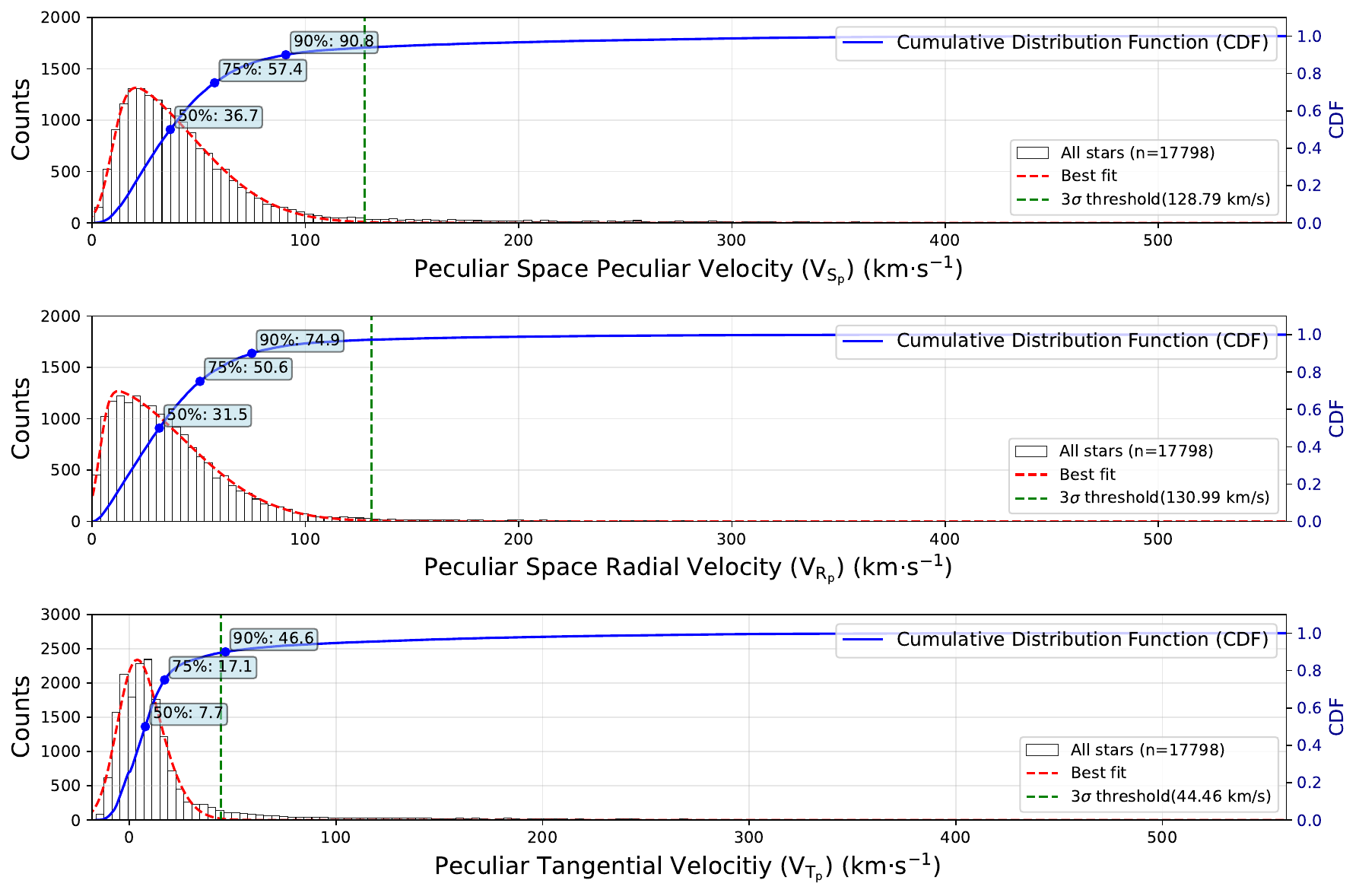}
    \caption{The three panels, from top to bottom, show the distributions of the peculiar space velocity, the absolute radial velocity, and the tangential velocity for the 17,798 OB stars. The red dashed line shows the skew-normal distribution fitted to the entire sample, while the blue line represents the empirical cumulative distribution. The 50th, 75th, and 90th percentiles of the sample are marked. The green dashed line indicates the 3$\sigma$ value derived from the skew-normal fit.} 
    \label{Figure07}
\end{figure*}
The Toomre diagram is widely used to distinguish thin-disk, thick-disk, and halo stars \citep{Nissen2010A&A...511L..10N,Bonaca2017ApJ...845..101B}. Figure~\ref{Figure06} shows the distribution of these 20,397 OB stars satisfying RUWE $\leq$1.4, $\sigma_\omega$/ $\omega\leq$30\%, and having LAMOST or \textit{Gaia} radial velocities in the Toomre diagram. The black dashed semicircle with V$_{\rm total}\geq$180\,km$\cdot$s$^{-1}$ and the orange dashed semicircle with V$_{\rm total}\geq$220\,km$\cdot$s$^{-1}$ represent the dividing lines between halo and disk stars, as defined by \citet{Nissen2010A&A...511L..10N} and \citet{Bonaca2017ApJ...845..101B}, respectively. Regardless of which dividing line is applied, most OB stars are located within the Galactic disk. Additionally, some stars exhibit relatively high spatial velocities, and a subset of these are classified as high-velocity star candidates. \citep{Hill1988Natur.331..687H,Bromley2006ApJ...653.1194B,Zheng2014ApJ...785L..23Z,Li2021ApJS..252....3L}.

\subsection{Identification of runaway OB stars}

To identify OB runaway stars in our sample, we calculate the peculiar space (V$_{\rm S_{\rm P}}$), radial (V$_{\rm R_{\rm P}}$), and tangential velocities (V$_{\rm T_{\rm P}}$) of 20,397 OB stars. Following the description of \citet{Wang2022ApJS..260...35W} and \citet{Guo2024ApJS..272...45G}, we first transformed the equatorial coordinates of the sample stars into the Galactic coordinate system, adopting the north Galactic pole coordinates ($\alpha_{G}=192\rlap{.}{^\circ}95948$, $\delta_{G}=27\rlap{.}{^\circ}12825$) from the \textit{Hipparcos} Consortium.

Subsequently, we performed a matrix transformation on the peculiar motions of the sample stars according to the equations listed by \citet{Moffat1998A&A...331..949M,Moffat1999A&A...345..321M}, and calculated both the tangential and radial peculiar velocities of the sample stars by subtracting the contributions of the solar motion and differential Galactic rotation. In this process, we adopted a solar motion of ($U_\odot$, $V_\odot$, $W_\odot$) = (11.10, 12.24, 7.25) km$\cdot$s$^{-1}$ from \citet{Schonrich2010MNRAS.403.1829S}, a solar Galactocentric distance $R_{\rm 0}$ = 8.127\,kpc, and a circular Galactic rotational velocity $V_{\rm c}$ = 229 km$\cdot$s$^{-1}$ from \citet{GRAVITYCollaboration2018A&A...615L..15G}. The circular velocity curve of the Milky Way is taken from \citet{Eilers2019ApJ...871..120E}. Runaway stars are generally identified based on the following three main criteria: 

(i) V$_{\rm S_{\rm P}}$ criterion:
\citet{Berger2001ApJ...555..364B} defined runaway stars as those with V$_{\rm S_{\rm P}}$$>$40\,km$\cdot$s$^{-1}$. and identified 23 Be stars using this cutoff based on the \textit{Hipparcos} proper motions and published radial velocities. \citet{Guo2024ApJS..272...45G} identified 229 runaway candidate stars in LAMOST DR8 data using V$_{\rm S_{\rm P}}$$>$43\,km$\cdot$s$^{-1}$. 

(ii) V$_{\rm R_{\rm P}}$ criterion: \citet{Cruz1974RMxAA...1..211C} and \citet{Vitrichenko1965IzKry..34..193V} identified Galactic OB runaway stars using $|V_{\rm R_{\rm P}}|$$>$30\,km$\cdot$s$^{-1}$. 

\begin{figure}[!htp]
    \centering
    \includegraphics[width=0.9\linewidth]{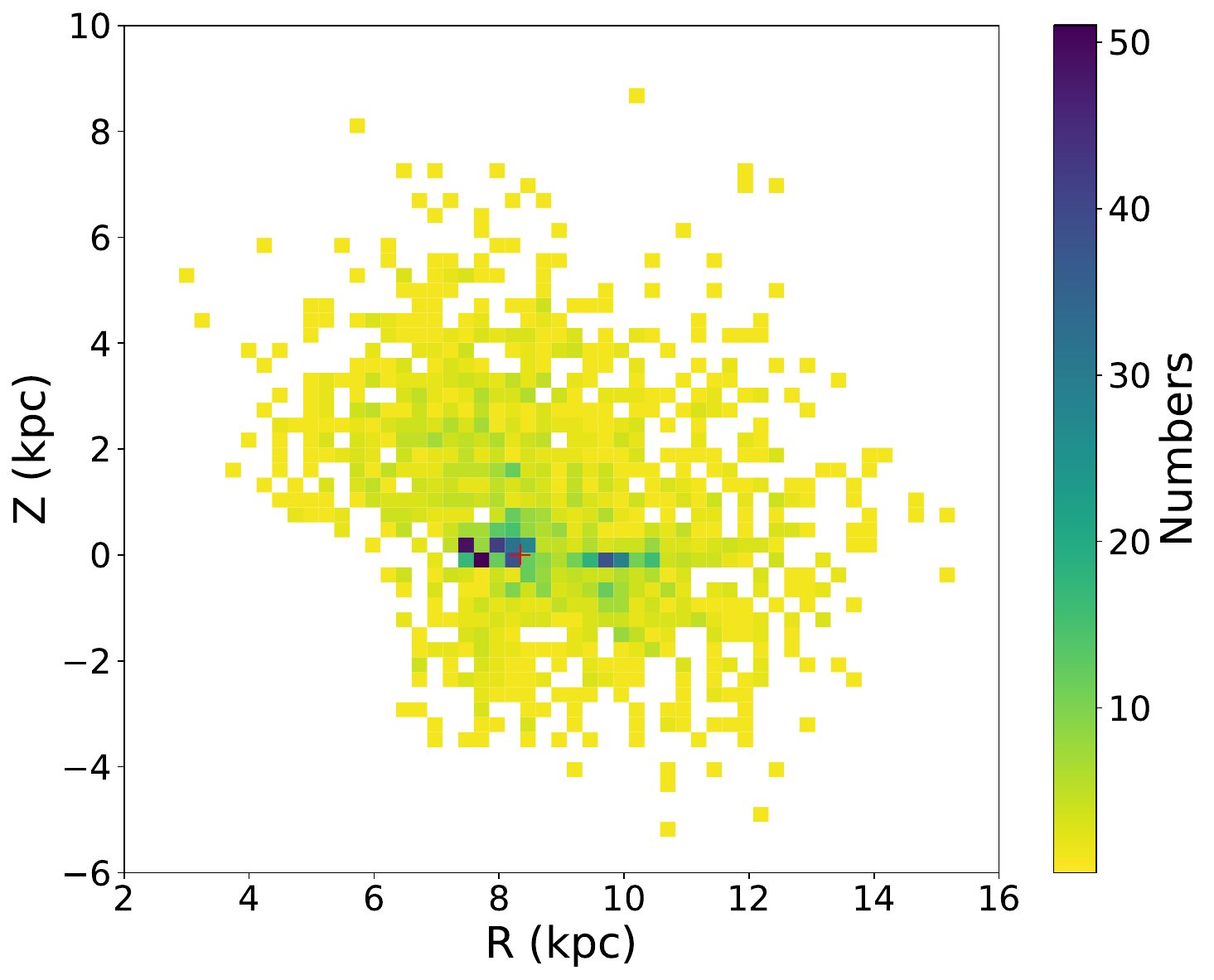}
    \caption{Distribution of the 1960 runaway star candidates in the Galactic R-Z plane. The color bar corresponds to the number density of stars. The red cross marks the solar position (R=8.34 kpc, Z=0 kpc) from \citet{Reid2014ApJ...783..130R}.} 
    \label{Figure08}
\end{figure}

(iii) V$_{\rm T_{\rm P}}$ criterion: \citet{Mdzinarishvili2005A&A...431L...1M} selected 61 runaway OB star candidates using V$_{\rm T_{\rm P}}$ $>$42\,km$\cdot$s$^{-1}$, based on a Monte Carlo method and \textit{Hipparcos} proper motions. \citet{Wang2022ApJS..260...35W} identified 16 Be runaway stars from LAMOST Medium Resolution Survey DR7 data by using V$_{\rm T_{\rm P}}$$>$24\,km$\cdot$s$^{-1}$

Considering the uncertainty of space velocity, we selected 17,798 stars satisfying $\frac{{\rm V_{\rm S_{\rm P}}}_{\rm err}}{\rm V_{\rm S_{\rm P}}}\leq$1, $\frac{{\rm V_{\rm S_{\rm R}}}_{\rm err}}{\rm V_{\rm S_{\rm R}}}\leq$1,  and $\frac{{\rm V_{\rm S_{\rm T}}}_{\rm err}}{\rm V_{\rm S_{\rm T}}}\leq$1 as our initial sample for further analysis. Figure~\ref{Figure07} displays the distributions of V$_{\rm S_{\rm P}}$, V$_{\rm R_{\rm P}}$, and V$_{\rm T_{\rm P}}$ for these 17,798 stars. Following the approach of \citet{Wang2022ApJS..260...35W} and \citet{Guo2024ApJS..272...45G}, and taking into account the characteristics of our data, we fitted a skew-normal distribution to each of the three velocity components, with the results overplotted in Figure~\ref{Figure07}. The top, middle, and bottom panels of Figure~\ref{Figure07} show that 90\% of the stars have V$_{\rm S_{\rm P}}<$90.8\,km$\cdot$s$^{-1}$ with a 3$\sigma$ threshold of 128.79\,km$\cdot$s$^{-1}$ for the fitted distribution, V$_{\rm R_{\rm P}}<$74.9\,km$\cdot$s$^{-1}$ with a 3$\sigma$ threshold of 130.99\,km$\cdot$s$^{-1}$ for the fitted distribution, and V$_{\rm T_{\rm P}}<$46.6\,km$\cdot$s$^{-1}$ with a 3$\sigma$ threshold of 44.46\,km$\cdot$s$^{-1}$ for the fitted distribution, respectively. We classify any star with a peculiar velocity exceeding its corresponding $3\sigma$ threshold (V$_{\rm S_{\rm P}}\geq$128.79\,km$\cdot$s$^{-1}$, V$_{\rm R_{\rm P}}\geq$74.9\,km$\cdot$s$^{-1}$, or V$_{\rm T_{\rm P}}\geq$44.46\,km$\cdot$s$^{-1}$) as a runaway candidate, resulting in a total of 1960 identifications. By cross-matching with the catalogs of \citet{Li2023AJ....166...12L,Guo2024ApJS..272...45G,Li2021ApJS..252....3L}, we find that 1949 of these are newly identified runaway star candidates. The complete list of runaway candidates is provided in Table~\ref{Table3}.

The spatial distribution of these 1960 runaway candidates in the Galactic R–Z plane is presented in Figure~\ref{Figure08}. The majority of these stars are located in the Galactic disk. There are also some stars with larger Galactic heights, which provide an important data basis for studying the origin of runaway stars \citep{Silva2011MNRAS.411.2596S, McEvoy2017ApJ...842...32M, Liu2023MNRAS.519..995L}.

\begin{table*}[!htp]
    \footnotesize
    \centering
    \caption{Information of 1960 OB runaway star candidates identified in LAMOST DR10. The parameters from left to right are: Obsid, Designation and equatorial coordinates (RA and Dec) from LAMOST, stellar name from \textit{Gaia} DR3, photogeometric distance (d$_{sun}$) from \citet{Baj2021AJ....161..147B}, peculiar space velocity (V$_{\rm S_P}$), peculiar radial velocity (V$_{\rm R_P}$), and peculiar tangential velocity (V$_{\rm T_P}$).}
    \begin{tabular}{rcllrrrrc}
    \hline
    \hline
    \multicolumn{1}{c}{Obsid} & Designation & \multicolumn{1}{c}{RA} & \multicolumn{1}{c}{Dec} &  \multicolumn{1}{c}{\textit{Gaia} DR3 Name} & \multicolumn{1}{c}{d$_{sun}$}&  \multicolumn{1}{c}{V$_{\rm S_P}$} &  \multicolumn{1}{c}{V$_{\rm R_P}$} & \multicolumn{1}{c}{V$_{\rm T_P}$}\\
    \cline{3-4}  \cline{7-9}
    \multicolumn{1}{c}{(LAMOST)}&\multicolumn{1}{c}{(LAMOST)}&\multicolumn{2}{c}{(LAMOST (deg))} &&\multicolumn{1}{c}{(pc)}&\multicolumn{3}{c}{(km$\cdot$s$^{-1}$)}\\
    \hline
   66604134 &J000009.65+115333.9 & 0.040234 & 11.892757 & 2765768108934958080 & 3869$\pm$442& 190.1$\pm$12.4 & -138.4 $\pm$3.8 & 130.3 $\pm$17.6\\
  370716216 &J000038.68+475126.5 & 0.161173 & 47.857362 & 387299363114015872 & 1872$\pm$76& 73.4$\pm$1.3 & -53.9$\pm$0.2 & 49.7$\pm$1.9\\
  281914160 &J000101.03+371709.1 &  0.2543324 & 37.285885 & 2880293927277360896 & 4438$\pm$325& 192.9 $\pm$4.1 & -168.2 $\pm$ 0.5 & 94.3 $\pm$ 8.4\\
  54904190 &J000106.49+172119.5 & 0.277053 & 17.355418 & 2773691047988421632 & 1669$\pm$71 & 168.7$\pm$ 1.2 & -156.6 $\pm$ 0.5 & 62.7 $\pm$ 3.1\\
  271004094 &J000210.76+420441.5 & 0.5448732 & 42.07821 & 384323229950201984 & 5731$\pm$739& 236.0 $\pm$ 6.5 & -213.3 $\pm$ 0.5 & 101.1 $\pm$ 15.1\\
  769112028 &J000220.66+064447.7 & 0.5861115 & 6.7465931 & 2745794346343664384 & 3289$\pm$ 153& 225.5 $\pm$ 12.7 & -110.7 $\pm$ 0.2 & 196.5 $\pm$ 14.6\\
   \nodata &  \nodata &  \nodata&  \nodata &  \nodata &  \nodata&  \nodata &  \nodata&  \nodata\\
   \nodata &  \nodata &  \nodata&  \nodata &  \nodata &  \nodata&  \nodata &  \nodata&  \nodata\\
   \nodata &  \nodata &  \nodata&  \nodata &  \nodata &  \nodata&  \nodata &  \nodata&  \nodata\\
    \hline
    \end{tabular}
    \vspace{0.3em}
    {\footnotesize 
    \raggedright 
    Note: This table is available in its entirety in a machine-readable form in the online journal.\\}
    \label{Table3}
\end{table*}

\section{Summary} \label{sec:Summary}

In this work, we identify 48,463 spectra corresponding to 34,550 OB stars from LAMOST DR10, based on their distributions in the H–R diagram constructed using \textit{Gaia} DR3 data and spectral line index spaces. Cross-matching with previous catalogs reveals that 6907 of these OB stars are newly identified.

Based on the MKCLASS code and LAMOST low-resolution spectra, we obtained the spectral types for 34,550 stars. The distribution of 22,413 OB stars with MKCLASS quality flags of "good" or better in different Galactic latitudes suggests that the stars located at low Galactic latitudes have more early-type OB stars. 

Combining \textit{Gaia} astrometric data, photogeometric distances from \citet{Baj2021AJ....161..147B}, LAMOST corrected radial velocities, and Gaia radial velocities, we calculate the spatial positions of 25,287 OB stars and three-dimensional velocities of 20,397 OB stars. Their distributions in both spatial position and the Toomre diagram indicate that they are mainly located in the Galactic disk. The newly identified OB stars show a spatial distribution similar to that of the previously known sample. 

Furthermore, based on the peculiar velocity distribution of 17,798 OB stars, we identify 1960 runaway star candidates using the threshold of V$_{\rm S_{\rm P}}\geq$128.79\,km$\cdot$s$^{-1}$, V$_{\rm R_{\rm P}}\geq$74.9\,km$\cdot$s$^{-1}$, or V$_{\rm T_{\rm P}}\geq$44.46\,km$\cdot$s$^{-1}$. Among these, 1949 are newly identified candidates.


\begin{acknowledgments}
We thank the anonymous referee for the helpful suggestions to help improve this manuscript. This study is supported by the National Natural Science Foundation of China under grants No. 12403034, 12573026; the Natural Science Foundation of Hebei Province A2024205031; Hebei Province Yan-zhao Golden Peak Talent Program (Postdoctoral Platform) for Key Talents under grant No. B2025003010, and the Science Foundation of Hebei Normal University (Nos. L2024B54, L2026B51 ,and L2024B56).  

This paper makes uses of data processed by the \textit{Gaia} Data Processing and Analysis Consortium (DPAC; \url{https://www.cosmos.esa.int/web/gaia/dpac/consortium}) and obtained by the \textit{Gaia}  mission from the European Space Agency (ESA) (\url{https://www.cosmos.esa.int/gaia}), 
This work made use of the data from LAMOST (Large Sky Area Multi-Object Fiber Spectroscopic Telescope, also known as the Guoshoujing Telescope) (\url{https://cstr.cn/31118.02.LAMOST}). LAMOST is a Chinese national mega-science facility, operated by National Astronomical Observatories, Chinese Academy of Sciences. Research has made use of the SIMBAD database, operated at CDS, Strasbourg, France.
\end{acknowledgments}

\appendix
\section{The flowchart of the OB star identification process in LAMOST DR10.}
\renewcommand{\thefigure}{A1}
\begin{figure*}[!htp]
    \centering
    \includegraphics[width=0.75\linewidth]{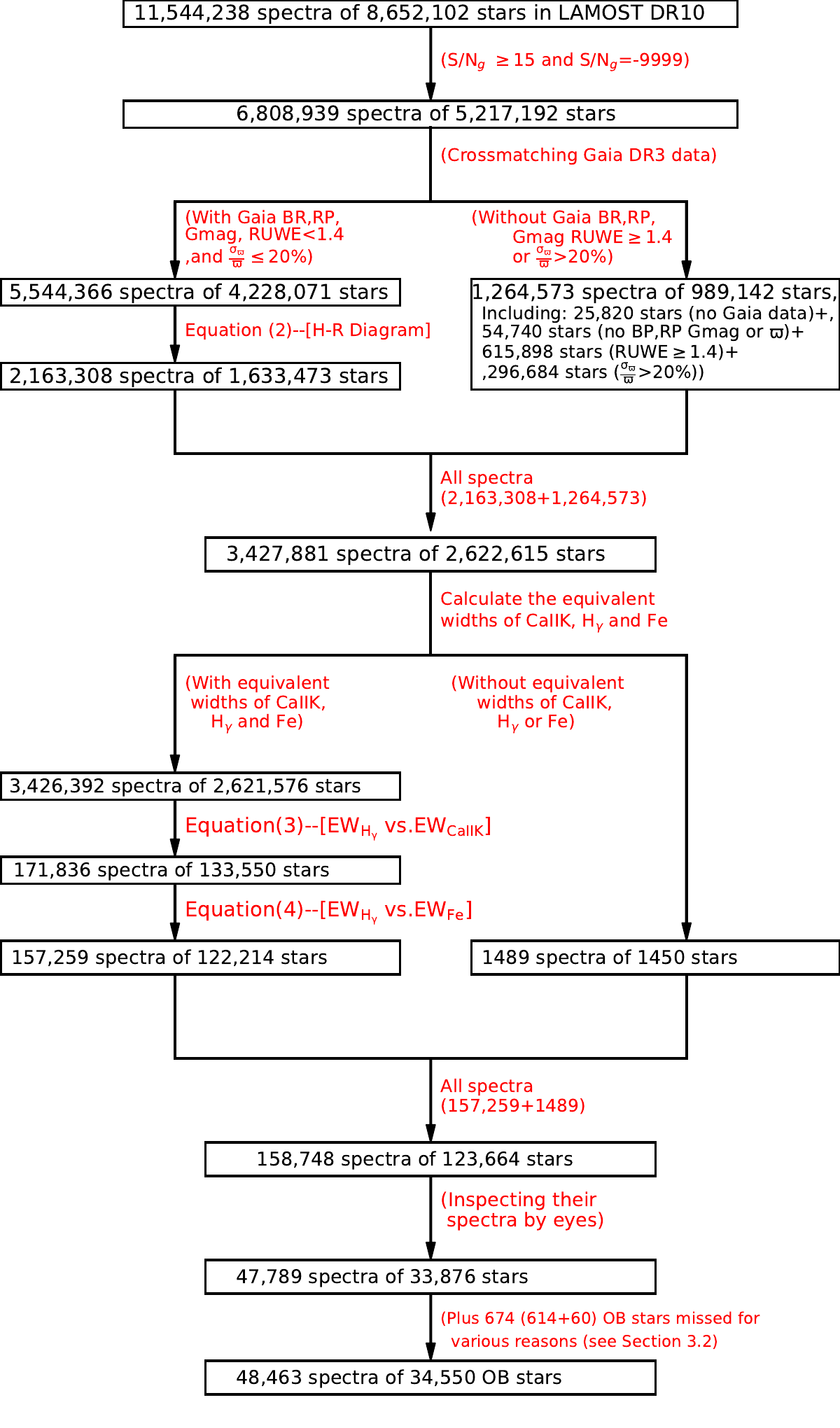}
    \caption{The flowchart of the OB star identification process in LAMOST DR10.}
    \label{figureA1}
\end{figure*}

\section{MKCLASS classification results of 34,550 OB-type stars}\label{MKOB}

\renewcommand{\thefigure}{B1}
\begin{figure}[!htp]
    \centering
    \includegraphics[width=1.0\linewidth]{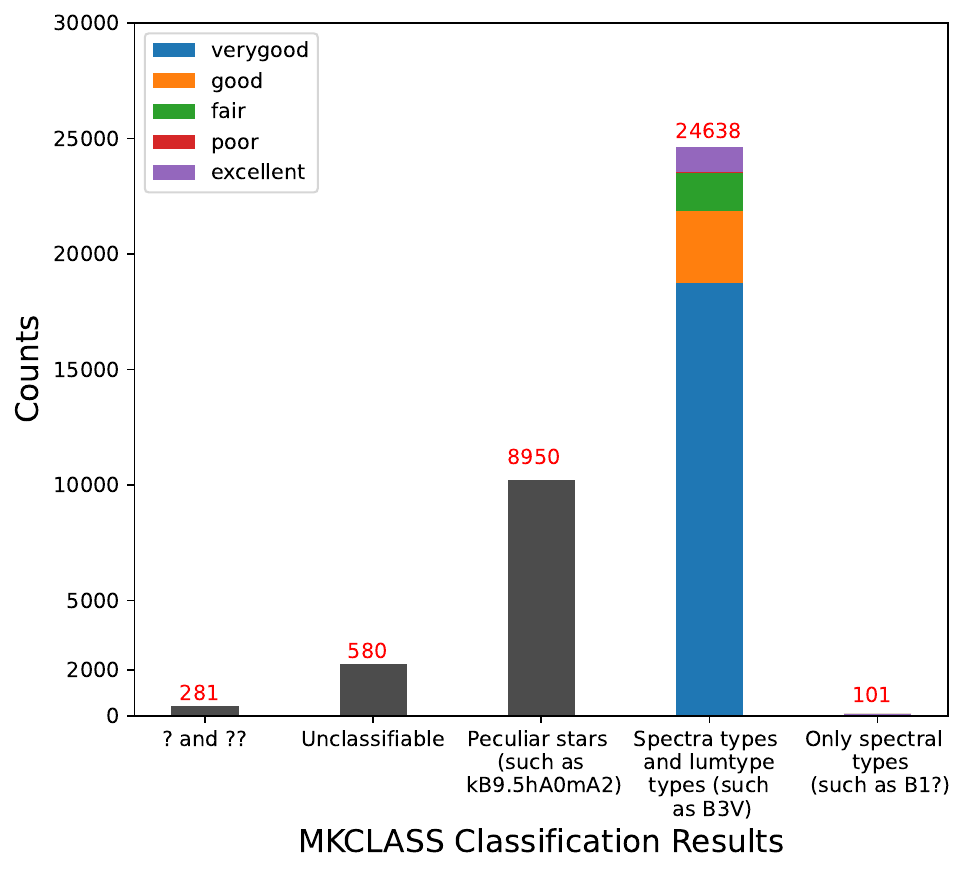}
    \caption{The histogram shows the distribution of MKCLASS classification results for 34,550 OB stars. The numbers of OB stars with different classification results are marked in the picture. The classification performance given by MKCLASS ("excellent", "very good", "good", "fair", and "poor") is indicated by different color bars.}\label{figGaia1}
\end{figure}

Figure~\ref{figGaia1} presents the distribution of MKCLASS classification results for 34,550 OB stars. Among these, 281 stars with labels of '?' or '??' and 580 stars labeled 'Unclassifiable' lack assigned spectral types. These unclassified stars mainly consist of Be-type stars, OB-type stars with low S/N$_{\rm g}$, and peculiar B-type stars. Additionally, for 8950 stars, a unique type cannot be determined from their characteristic lines; instead, they show discrepant classifications, such as "kB9.5hA0mA2", indicating that the K-line-based type is B9.5, the hydrogen-line-based type is A0, and the metallic-line-based type is A2. Furthermore, 101 stars are assigned only spectral types (without luminosity classes), while 24,638 stars are assigned both spectral types and luminosity classes. This suggests that LAMOST survey data includes many peculiar late B-type stars \citep{Gray2009ssc..book.....G,Xiang2022A&A...662A..66X,Liu2024ApJS..275...24L}.

In Figure~\ref{figGaia2}, we show the distribution of 22,960 out of 24,638 OB stars with MKCLASS quality evaluations of “good” or better in the spectral subtype versus luminosity class plane. Most of the stars have spectral types from B0 to A0 and luminosity classes from III to V. The highest stellar density occurs in the late-B to early-A region (B7–A1), particularly for luminosity classes II–V, whereas O-type and early-B stars are considerably rarer. B-type stars with spectral types earlier than B3 are rare, which is due to the observational selection effects of the LAMOST survey. In addition, there are a small number of stars that are misclassified as F-type stars due to their low spectral S/N$_{\rm g}$.

\renewcommand{\thefigure}{B2}
\begin{figure*}[!t]
    \centering
    \includegraphics[width=1.0\linewidth]{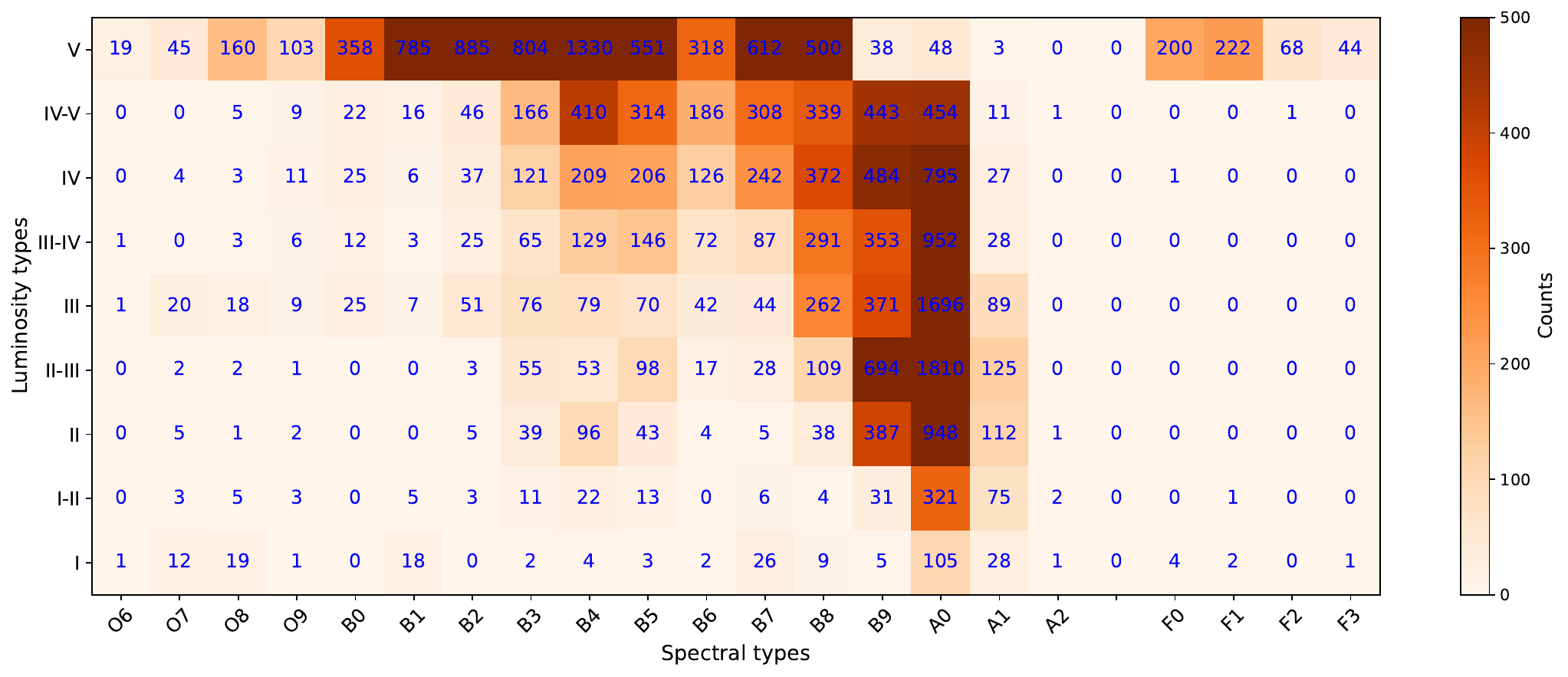}
    \caption{Distribution of 22,960 OB stars whose MKCLASS quality assessments are better than “good” and including “good” in the spectral subtype vs. luminosity class plane. The numbers in the different color boxes represent the number of stars in this spectral-type range.}
    \label{figGaia2}
\end{figure*}

\bibliography{sample701}{}
\bibliographystyle{aasjournalv7}



\end{document}